\documentclass[]{tMOP2e}

\begin{document}

\title{A scaling law for light scattering from dense and cold atomic ensembles}
\author{I.M. Sokolov${}^{1,2}$, A.S. Kuraptsev${}^{1}$, D.V. Kupriyanov${}^{1}$, M.D. Havey${}^{3}$, and S. Balik${}^{3}$\\
\small $^{1}$Department of Theoretical Physics, State Polytechnic
University, 195251, St.-Petersburg, Russia \\ \small $^{2}$Institute for
Analytical Instrumentation, RAS, 198103, St.-Petersburg, Russia  \\\small$^{3}$Department of Physics, Old Dominion University,
Norfolk, VA 23529}

\date{\today }
\maketitle

\begin{abstract}
We calculate the differential cross section of polarized light scattering from a cold and dense atomic ensemble. The regularities in the transformation of the cross section when increasing the size of the atomic ensemble are analyzed numerically. We show that for typical experimental conditions, an approximate scaling law can be obtained.  Very good agreement is found in a comparison with experimental data on the size dependence of a dense and cold cloud of $^{87}$Rb atoms.

\end{abstract}

\section{Introduction}
Recent interest in cold, but not quantum degenerate, atomic gases with high atomic densities is stimulated in part by their unique physical properties as well as by their wide possible applications in different areas of quantum electronics. For instance, such gases can be used in quantum information science, in which cold clouds are promising objects for quantum memory problems \cite{PSH,Simon,Lam,GSOH,MSLOFSBKLG}. Theoretical results have also suggested that a high degree of vacuum squeezing of near resonance radiation may be obtained in such optically dense media through the action of the polarization self rotation effect \cite{Matsko,Mikhailov,Horrom}.  They have also been considered as a possible basis in which to explore the possibility of development of an atomic physics based random laser \cite{Guerin}.
	
In these applications, and in many others, interaction of the cold atoms with a quasi resonant electromagnetic field plays a key role. This interaction in the case of dense clouds has some peculiarities in comparison with the more common situation where dilute gases are of interest.  In the latter case different atoms scatter light independently and we can consider light propagation inside the medium as a sequence of light scattering events from free atoms.  By dividing all scattering events into incoherent and coherent scattering (scattering into an initial mode) and describing coherent forward scattering by introduction of an effective atomic susceptibility we can theoretically treat light interaction with dilute cold clouds for most realistic experimental situation. This approach allows us to describe even such subtle phenomena as weak localization and coherent backscattering with a very good accuracy (see, for examples, the following reviews \cite{KH06}-\cite{MD}).

For more dense clouds, when the mean free path of a photon becomes comparable with the wave length of light in the medium, the resonant dipole-dipole interatomic interaction strongly interferes with the scattering process. The atomic dipoles cannot be considered as independent secondary sources of the scattered waves. Here we can envision dealing with light scattering from a huge quasi molecule consisting of all the atoms in the cloud. A consistent quantum description of such a process can be performed only for a relatively small ensemble. Computer simulations allow consideration of atomic clouds with several thousands of atoms. However, in a typical experiment the total number of atoms in a sample can be $10^{2}$ to $10^{3}$ greater.  Furthermore, such experiments require realistic modeling in order to extract the essential physics of the observed processes.  Theoretical approaches alternative to currently developed ones, even if approximate in nature, are then desirable.

One of such approaches is discussed in \cite{FKSH11}.  Because the problem of a dense atomic ensemble belongs to the field of electrodynamics of a continuum, it was suggested in \cite{FKSH11} to use a quantum microscopic calculation to obtain the dielectric susceptibility of such ensembles and to use it further in equations of macroscopic electrodynamics. In the present paper we discuss another possibility. Here we numerically analyze the differential cross sections for quasi resonant light scattering from dense atomic ensembles having different sizes, and so determine an approximate scaling law for the cross-section. The sequential deduction of a scaling law for three dimensional clouds of arbitrary but high density with a heterogeneous spatial distribution of atoms seems to be an extremely complex problem.  It is especially difficult for the case when the vector nature of the scattered radiation is important. Although this is so, at the same time we will show here that for some particular, but typical and practically important, experimental cases an approximate scaling law can be obtained.  For this purpose we analyze numerically the regularity in the transformation of the spectrum of the differential cross section when increasing the size of a dense and cold atomic ensemble. On the basis of this analysis we derive an approximate but quite simple scaling law.  The scaling agrees with heuristic arguments, which provide some useful physical picture of the overall processes.  The obtained results are compared with the limited experimental data currently available, with realistic attention paid to the characteristics of the atomic sample, and the unusual geometrical arrangement of the incident and scattered fields.

\section{Basic assumptions and approaches}
Development of a scaling law in this work will be performed on the basis of a consistent quantum microscopic approach.  We are interested in weak-field light scattering. In such a case we can use a method first suggested in 1964 by Stephan \cite{St64}, and D. A. Hutchinson and H. F. Hameka \cite{HH64}. In \cite{St64} and \cite{HH64} this approach was applied for description of collective effects in two-atom systems. Later this method was further generalized to the multi atom ensembles in numerous papers \cite{Lehmberg70}-\cite{SKH11}.

The basic idea of the applied approach is the following. We consider single photon scattering from an ensemble of motionless atoms which all are initially in the ground state.  The method is based on subsequent calculation of the many particle Green's function expressed by the resolvent operator, see below. The calculation of the resolvent operator is simplified by taking into consideration the main contributions to its self-energy part. Physically that assumes keeping, in the self-energy part, only the virtual states consisting of one excited atom and no photon, one photon and no excited atoms and two excited atoms and one photon. The latter contribution is off resonant and mostly responsible for the near-field longitudinal interaction at short inter-atomic distances.

The matrix size of the set of equations, defining the resolvent operator, is determined by the energetic structure of the atoms under study. The degeneracy of the atomic states essentially limits the size of the ensemble which can be calculated by this method. That clarifies why most of the studies of the cooperative effects in dense atomic systems were made for two level model systems. In the present paper we consider the simplest non-trivial energy configuration compromised with the vector nature of the electromagnetic field i. e. the case of four atomic states with a $J=0\rightarrow J=1$ transition.

The differential cross section of the photon scattering from an initial mode $\omega$ with wave vector $\mathbf{k}$ and polarization unit vector $\mathbf{e}$ scattering into an arbitrary final mode $\omega',\mathbf{k}^{\prime }\mathbf{,e}^{\prime } $ is given by
\begin{equation}
\frac{d\sigma }{d\Omega}=\frac{\omega'^{3}\omega}{c^{4}}\left\vert \sum\limits_{e,e^{\prime }}\left(
\mathbf{e}^{\prime \ast }\mathbf{d}_{g;e}\right) R_{ee^{\prime }}(\omega)\left(
\mathbf{ed}_{e^{\prime };g}\right) \exp \left( i(\mathbf{kr}_{e^{\prime }}-\mathbf{k}^{\prime
}\mathbf{r}_{e}\right) \right\vert ^{2}.  \label{1}
\end{equation}

Here the symbol $g$ denotes the ground state of the initial non disturbed atomic ensemble;  $\mathbf{d}_{g;e}$ is the matrix element of a dipole moment operator of that atom which is excited in the state $e$. The sum in (\ref{1}) is calculated over all possible single atom excited states.

The matrix $R_{ee^{\prime }}$ which is a part of (\ref{1}), plays a key role in the theory of cooperative scattering of light from a dense atomic ensemble. As was shown in \cite{SKKH09} and \cite{SKH11} this matrix is given by the projection of the resolvent operator of the jointly considered system on the states consisting of a single atom excitation, distributed over the ensemble, and the vacuum state for all the field modes. For the dipole-type long-wavelength approximation of interaction between atoms and field this projected resolvent can be found as the inverse matrix of the following expression:
\begin{equation}
R_{ee^{\prime }}(\omega )=\left[ \hbar(\omega -\omega _{0})\delta_{ee'}-\Sigma
_{ee^{\prime }}(\omega )\right] ^{-1}.  \label{2}
\end{equation}

The last term in the square brackets in the right-hand part of this expression is a self-energy part.  For the case when $e$  and $e'$ correspond to excitation of different atoms $\Sigma_{ee^{\prime }}$ describes excitation exchange between a pair of atoms inside the ensemble. While calculating the self-energy matrix we apply the standard pole or "on-shell" approximation, similar to the Wigner-Weiskopf approach for single atom spontaneous emission. In such a case $\Sigma _{ee^{\prime }}(\omega)\approx\Sigma _{ee^{\prime }}(\omega_0 )$ and
\begin{eqnarray}
\Sigma _{ee^{\prime }}(\omega_0 ) &=&\sum\limits_{\mu ,\nu  }\frac{\mathbf{d}_{e;g}^{\mu }\mathbf{d}_{g;e'}^{\nu }}{r^{3}}\left[ \delta _{\mu \nu }\left(
1-i\frac{\omega _{0}r}{c}-\left( \frac{\omega _{0}r}{c}\right) ^{2}\right)
\exp \left( i\frac{\omega _{0}r}{c}\right) \right. + \notag \\
&&\left. -\dfrac{\mathbf{r}_{\mu }\mathbf{r}_{\nu }}{r^{2}}\left( 3-3i\frac{\omega _{0}r}{c}-\left( \frac{\omega _{0}r}{c}\right) ^{2}\right) \exp
\left( i\frac{\omega _{0}r}{c}\right) \right] .\label{3}
\end{eqnarray}
Here $\mathbf{r}$ is the relative separation of two atoms excited in states $e$ and $e'$ and $\omega _{0}$ is the transition frequency of the atoms.

For the case when the state specifications $e$  and $e'$ belong to two different excited states of the same atom, the matrix elements are performed by the natural linewidth  $\gamma$ of the excited state for an isolated atom
\begin{equation}
\Sigma _{ee'}=\delta_{ee'}\left[-\frac{i\hbar \gamma}{2}+\hbar\Delta_L\right].
\label{4}
\end{equation}
where $\Delta_L$ is the non-converging term associated with the radiative Lamb-shift. We will further assume that this shift is included in the physical value of the transition frequency $\omega_0$.

With explicit analytical expressions (\ref{3}) - (\ref{4}) the resolvent $R_{ee^{\prime }}(\omega )$ can be numerically evaluated for an atomic system consisting of a macroscopic number of atoms. This allows us to determine both the angular and spectral dependence of the cross section (\ref{1}). In the following section we will analyze these dependencies for atomic clouds with different densities and different sizes.

\section{Results and discussion}
Relations (\ref{1}) - (\ref{4}) allow us to analyze the differential cross section for a wide range of parameters. We, however, will focus our attention on the area of characteristics typical for experiment (see, for example \cite{Archive}). In these experiments high density clouds are prepared in a quasi static electric dipole trap. The atomic spatial distribution for samples formed in such  traps can be depicted as follows:
\begin{eqnarray}
n(\mathbf{r})=n_0\,\exp\left(-\frac{x^2+z^2}{2r^2_{tr}}-\frac{y^2}{2r^2_l}\right).
\label{5}
\end{eqnarray}
Here $r_{l}$ and $r_{tr}$ are the longitudinal and transverse Gaussian radii of the cloud; $n_0$ is the peak density of the atoms at the center of the ellipsoidal sample.

In experiments with $^{87}$Rb, the peak density changes from relatively small values  $n_0\simeq 6\cdot\,10^{10}\,cm^{-3}$ corresponded to dilute ensemble where collective effects can be neglected to $n_0\simeq 5\cdot\,10^{13}\,cm^{-3}$. In the latter case the mean free path of quasi resonant photons becomes comparable with the averaged interatomic separations. To have possibilities to compare results of the theory with these experiments, we further choose the density of our motionless four levels atoms in such a way that photons would have the same mean free path as in the $^{87}$Rb samples. Estimating the resonant cross section of the light from a single atom with $J=0\rightarrow J=1$ transition as $3\lambda^2/2\pi$ we obtain that $n_0\simeq 5\cdot\,10^{13}\,cm^{-3}$ in experiment corresponds to $n_0\simeq 0.05\,k_0^{-3}$ in theory. Here $k_0$ is the wave number of the scattered light. Hereafter in this paper we will use $k_0^{-1}$ as the unit of length.

Another peculiarity of current experiments with cold clouds considered here is in the specific geometry of scattered light observation. Because of constraints on the vacuum chamber geometry, the probe light is directed at an angle of 30 degrees away from the fluorescence collection direction. The detected light propagates along z axis. In numerical simulations of the theory we will assume the same geometry.

Figure 1 shows the spectral dependence of light scattering for the clouds with different densities. Here we consider probe light linearly polarized perpendicularly to the plane of incidence. Detection without polarization analysis is assumed; this is in accordance with experiments done so far.
The graphs shown are obtained as result of averaging over more than $10^{3}$ realizations of random atomic positions.
\begin{figure}[th]
\begin{center}
{$\scalebox{0.8}{\includegraphics*{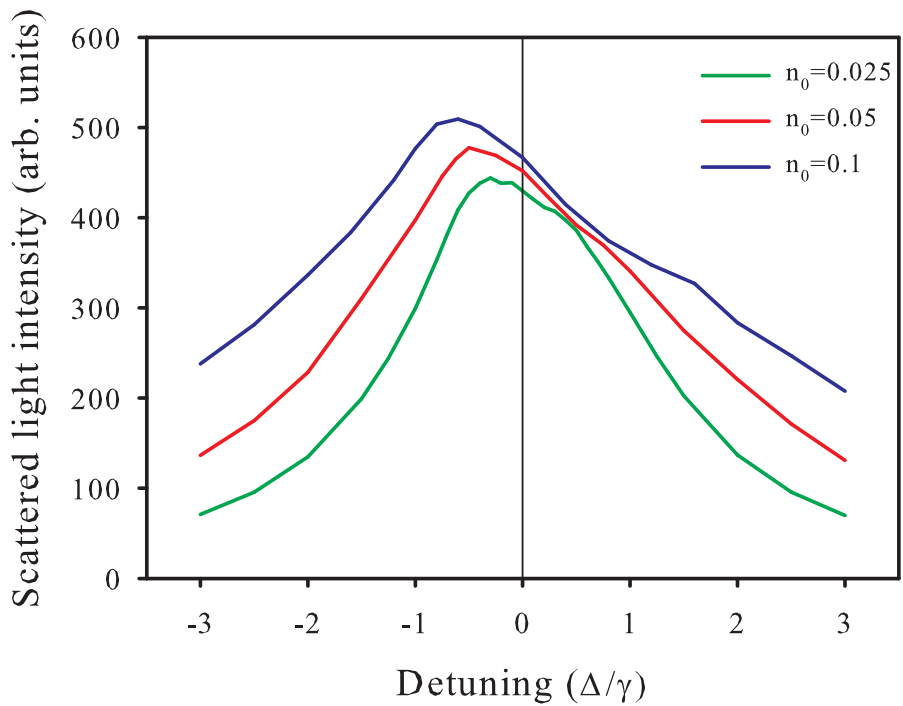}}$ }
\caption{ Intensity of light scattered from atomic clouds.
The angle of incidence is $\theta=\pi/6$, $\varphi=0$; the scattering angle is 0 (along the z axis).
Light is linearly polarized along x-axis. The spatially inhomogeneous atomic cloud is taken to be Gaussian with $r_{tr}=10$; $r_{l}=30$.}
\end{center}
\par
\label{fig1}
\end{figure}

The increasing width of the spectrum shown in Fig. 1 has two main causes. The first of these is the increasing of the optical depth of the cloud. For a spatially inhomogeneous atomic ensemble it is determined by the size of the system and the single atom scattering cross section $\sigma$ (for a given size of the system and for $n_0=0.1$ the optical depth along the z direction is $b_{tr}=\sqrt{2\pi}n_0r_{tr}\sigma\simeq 47$). A second reason for spectral broadening is the resonant dipole-dipole interatomic interaction. This interaction has an influence particularly on the observed red shift of the maximum of the spectrum as well as on distortion of its shape.  Due to the cooperative effects the profile differs noticeably from a Lorentz one even for the lowest considered density $n_0=0.025$.

\begin{figure}[th]
\begin{center}
{$\scalebox{0.8}{\includegraphics*{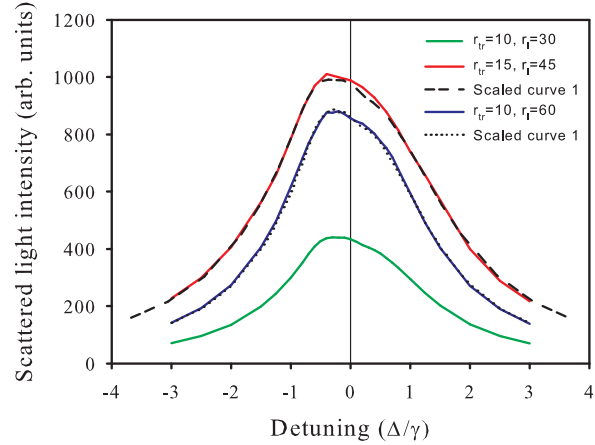}}$ }
\caption{ Intensity of light scattered from atomic clouds of different sizes; $n_0=0.025$.
Other conditions are the same as in Fig. 1}
\end{center}
\par
\label{fig2}
\end{figure}

As was mentioned above, the main purpose of this paper is to analyze how the curves from Fig. 1 transform when the size of the scattering cloud increases. Fig. 2 shows such a transformation for ensembles with $n_0=0.025$.  We give two examples.  First,
we examine the cloud with the same optical thickness along the direction of the scattered light $b_{tr}$ but with a larger cross sectional area. The second calculation corresponds to the case when we increased all three Gaussian radii of the cold ensemble. Increasing of the sizes caused both increasing of amplitude and broadening of the profile. But the shape of the spectrum for a given density can be obtain from the initial one by scaling. The intensity of light scattered to a small angle, nearly in the forward direction obviously should depend on the cross section area of scattered cloud. For a specified optical depth along this direction this dependence is very close to linear. To understand the role of optical depth let us take into account that different detected photons undergo scattering of different order inside the medium. The contribution of single scattering comes mainly from atoms which are located in the dilute boundary region of the cloud where the transverse optical thickness is less or comparable with unity. As detuning increases the number of such atoms increases too, but the probability of the scattering decreases.  These two mechanisms compensate each other approximately. This compensation takes place only until $b_{tr}$ is more than unity. As $b_{tr}$ becomes equal to unity, the single scattering contribution falls to zero very sharply.  We know that for not very dense clouds $b_{tr}$ decreases with detuning $\Delta$ from exact resonance approximately as $b_{tr}(\Delta)=b_{tr}(0) \frac{\gamma^2/4}{\Delta^2+\gamma^2/4}$, so the "boundary" value $b_{tr}=1$ is reached approximately (for large $b_{tr}(0)$) for $\Delta\sim\gamma\sqrt{b_{tr}(0)/4}$. This means that the spectral width of partial single scattering contribution is proportional to $\sqrt{b_{tr}(0)}$. A similar estimation can be performed for scattering of higher order.

These things together then suggest the following scaling law for the spectrum of light scattered from a dense cloud
\begin{eqnarray}
I(r_{x_1},r_{y_1},r_{z_1},\Delta)=I\left(r_{x_0},r_{y_0},r_{z_0},\Delta\sqrt{\frac{r_{z_0}}{r_{z_1}}}\right)\frac{r_{x_1}}{r_{x_0}}\frac{r_{y_1}}{r_{y_0}}.
\label{6}
\end{eqnarray}
Here $r_{x_i},\,r_{y_i}, \,r_{z_i}$ for $i=0,1$ are three Gaussian radii for two different atomic clouds with the same peak density $n_0$.

Expression (\ref{6}) allows us to predict the results of light scattering from a large atomic ensemble through knowing the corresponding result for a smaller one. In Fig. 2 we compare scattering spectra obtained in quantum calculation (solid lines) with prediction of Eq. (\ref{6}) (dashed lines). The coincidence of the two is well within the range of accuracy of the calculation.

Fig. 2 depicts the total intensity of scattered light, that is, the intensity summed over all possible polarizations. Fig. 3 shows that the suggested scaling law is valid also for different polarization channels separately. In Fig. 3a we consider scattering of light linearly polarized perpendicularly to the plane of incidence (so called $s$ polarization). Two curves correspond to two final polarization of the light $s$ and $p$. As above, in dashed lines scaled results are shown. Note that the polarization degree of scattered light is significant. This suggests the relatively large contribution of scattering of low orders. But in spite of the large role of lower order scattering, which takes place in the dilute boundary region of the Gaussian ensemble, the density effects on such a shift and profile distortion are also significant.

\begin{figure}[th]
\begin{center}
{$\scalebox{0.45}{\includegraphics*{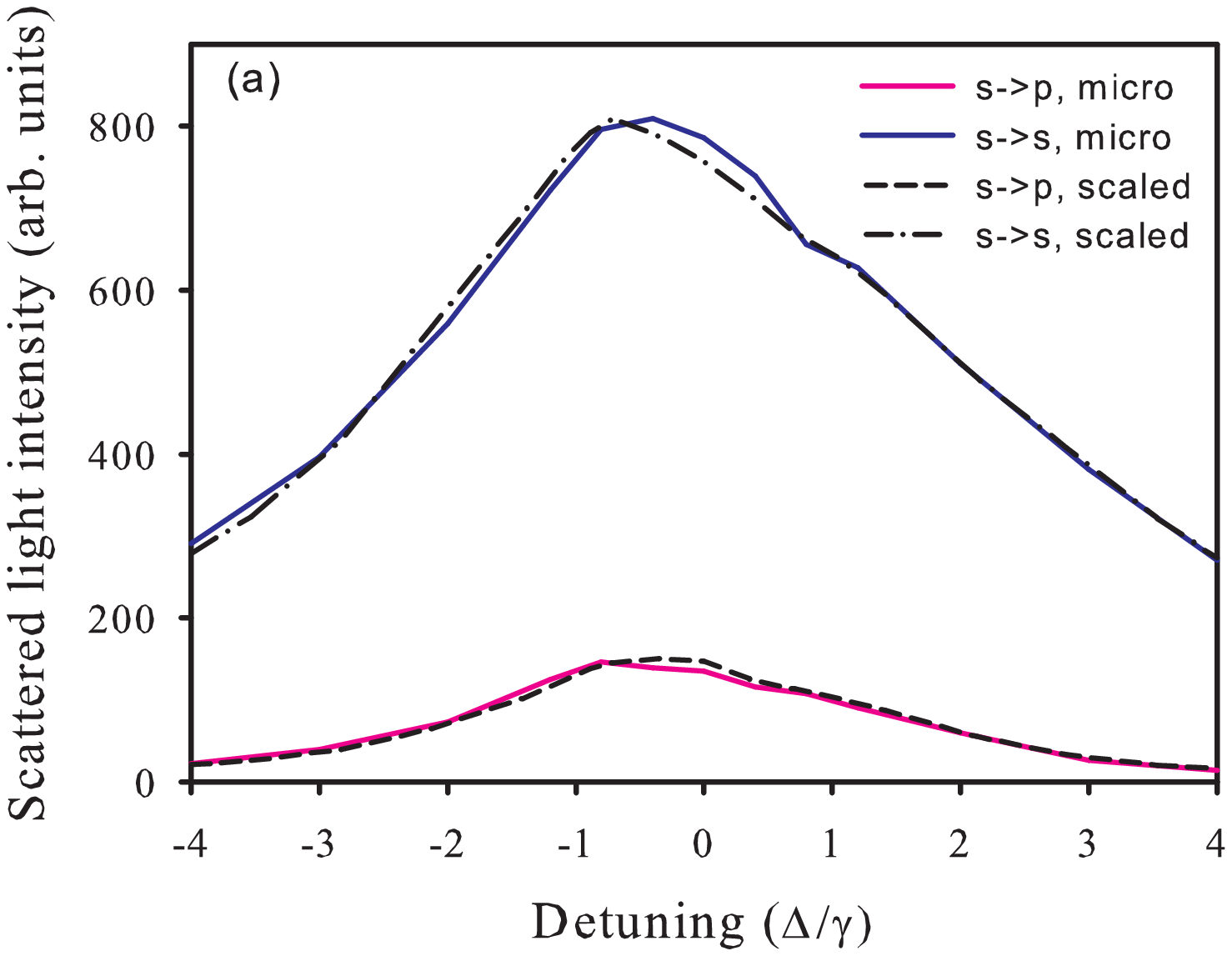}}$ }
{$\scalebox{0.45}{\includegraphics*{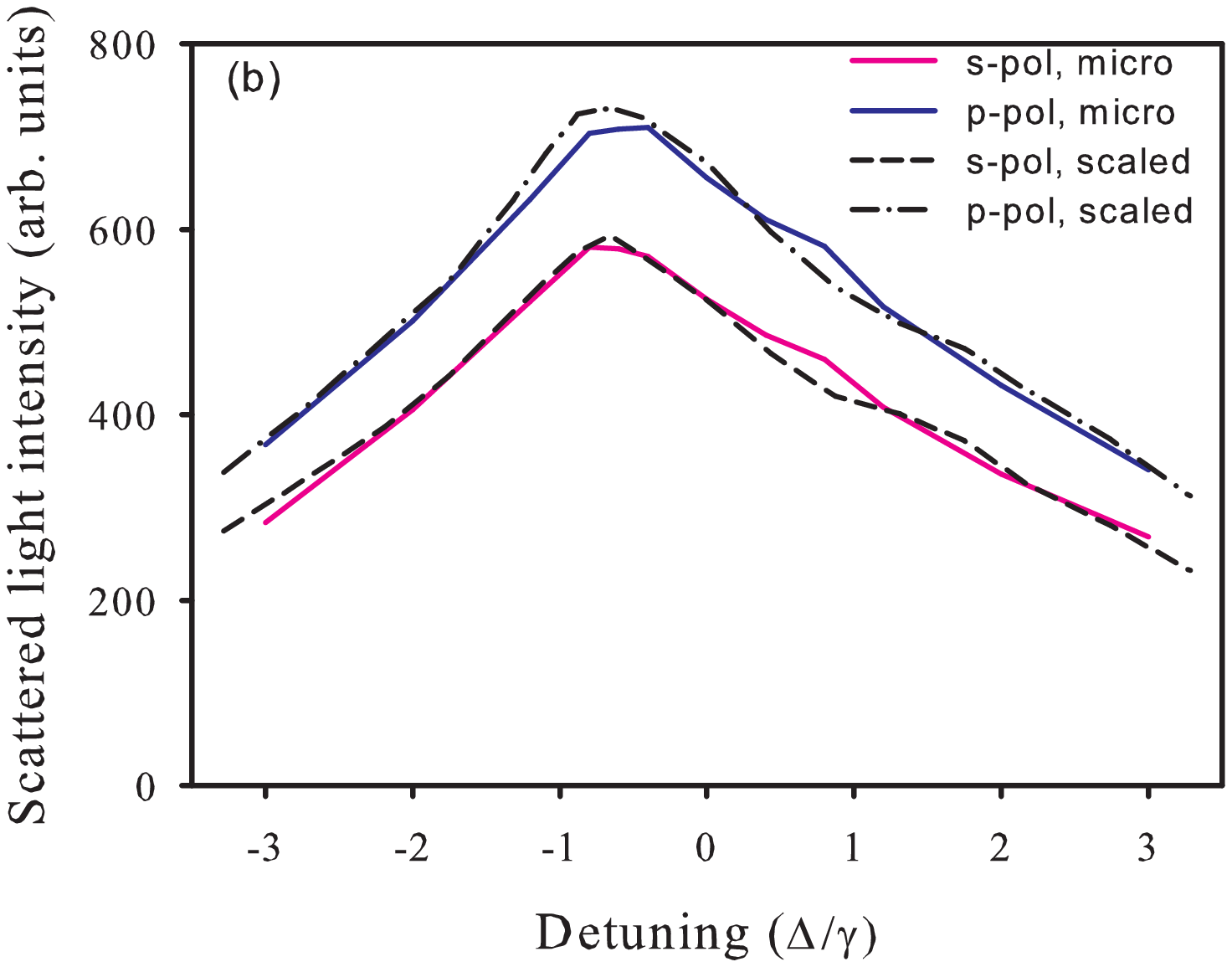}}$ }
\caption{ Polarization components of light scattered from atomic clouds of different sizes; (a)$n_0=0.05$, $r_{tr}=20$, $r_l=30$; (b) $n_0=0.1$, $r_{tr}=12$, $r_l=36$. Dashed lines obtained with Eq. (\ref{6}) by means of scaling of dates calculated for clouds with $r_{tr}=10$, $r_l=30$ and corresponding densities. }
\end{center}
\par
\label{fig3}
\end{figure}

Fig. 3b substantiates the scaling law (\ref{6}) for light with different initial polarizations. Here we show the total scattered light intensity for $s$ and $p$ linearly polarized probe beams. Note that Fig. 3a and Fig. 3b are considered for clouds with $n_0=0.05$ and $n_0=0.1$ respectively. These are relatively large densities. As was mentioned previously, the former corresponds to $n_0\simeq 5\cdot\,10^{13}\,cm^{-3}$ for $^{87}$Rb which is very close to the ultimate accessible density in experiments so far. For density $n_0=0.1$ the considered cloud has an optical depth equal approximately to 56. It is a very large value and is close to that can be achieved in real experiments. Good agreement between results obtained in self consistent calculations (solid lines in Fig. 2 - 3) and scaled ones (dashed lines) allows us to make the conclusion that the discussed scaling (\ref{6}) can be extrapolated to larger ensembles and can be used for description of future experiments.

The coincidence of scaling and the calculated curves in Figs. 3a and 3b is evidently not quite as good as that seen in Fig. 2. We note, however, that the considered cases correspond to large clouds with large number of atoms. The calculation for such ensembles is very time consuming, and the results shown in Figs. 3a and 3b are obtained by averaging over fewer configurations, 400 and 250 clouds correspondingly. So in this case, the coincidence of calculated and scaled curves in Figs. 3a and 3b are also within the range of accuracy of the calculation.

Figs. 2 - 3 together establish the applicability of the obtained scaling law for small-angle scattering. It is clear that the qualitative reasonings given above in the writing of Eq. (\ref{6}) are not valid for the scattering at large angles. So it is important to understand the applicable range of utility of expression (\ref{6}). To determine the limitations of expression (\ref{6}) we calculate the angular distribution of scattered light and its change with an increase in the dimensions of the scattering medium (see Fig. 4).

In Fig. 4, with a solid line we show the spatial distribution of scattered light for two polarization channels. To illustrate the range of phenomena, we consider here right handed circularly polarized probe light and right and left circularly polarized scattered light. Probe light falls at an angle of $\pi/6$ to the axis of the cigar-shaped cloud. The incident light is in exact resonance with the free atom transition. The number of Monte-Carlo tests is 3000 for small clouds and 1700 for larger ones. We did not smooth obtained results. The residual scatter in the curves indicates the accuracy of our calculation.

The calculated curves reproduce all the expected characteristics of light scattering. First, we see the main diffraction peak. There is also a suggestion of  subsidiary maxima. Here we deal with diffraction on a spatially inhomogeneous object, which is why the diffraction feature has such a specific appearance.  In both polarization channels in Fig.4 one also sees a coherent back scattering cone. There is also a contribution of incoherently scattered light.

\begin{figure}[th]
\begin{center}
{$\scalebox{0.8}{\includegraphics*{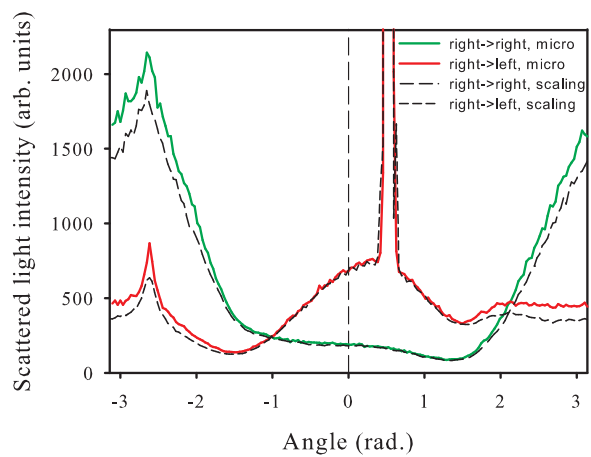}}$ }
\caption{Angular distribution of right circularly polarized light scattered from atomic clouds; The angle of incidence is
$\theta=p/6$, $\varphi=0$; $r_{tr}=15$, $r_l=45$; $n_0=0.025$.
In solid lines we show results of Monte-Carlo simulations. Dashed curves represent results calculated by means of scaling of the angular distribution obtained for a cloud with $r_{tr}=10$, $r_l=30$; $n_0=0.025$}
\end{center}
\par
\label{fig5}
\end{figure}

In the dashed lines in Fig. 4 we show the approximate angular distribution calculated by means of the scaling law on the basis of the corresponding distribution for a small ensemble. For exact resonance radiation ($\Delta=0$) it means a simple multiplication of scattered light intensity by the ratio of cross section areas of two clouds.

The scaling law evidently does not describe coherent effects. For example, it does not reproduce the amplitude of the principal diffraction peak. Divergences with the calculation of the coherent backscattering cone are also visible. At the same time diffusive scattering, especially that part which propagates into the front half-space (with respect to incident light), is described sufficiently accurately. As far as scattering into the rear half-space is concerned, in the presence  of the definite quantitative divergences scaling reproduces the basic qualitative regularities sufficiently well. And this scaling law can be used for the qualitative analysis of the results of experiments even if these experiments study backwards scattering.

For near forward scattering some aspects of the scaling law were examined in experiment. We have for example checked the dependence of the intensity of scattered resonant light on optical depth of the atomic ensemble at constant atom number (see Fig. 5).
\begin{figure}[th]
\begin{center}
{$\scalebox{0.8}{\includegraphics*{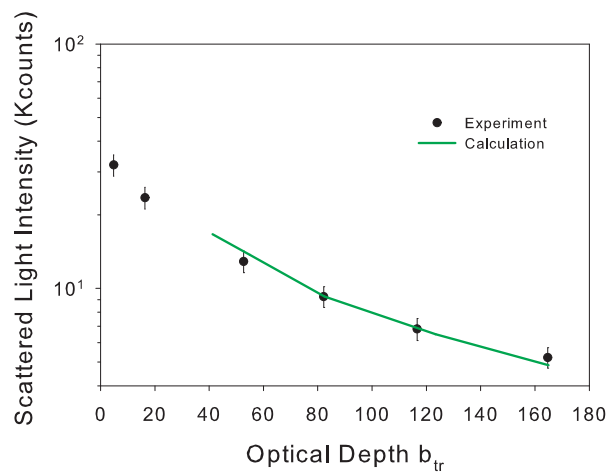}}$ }
\caption{Comparison of theoretical and experimental results for on-resonance excitation of a dense atomic cloud.  The light is collected with the same excitation-detection geometry as referred to in Figs. 1 - 3.  The optical depth of the cloud is varied by allowing it to expand while conserving the total number of atoms in the sample.}
\end{center}
\par
\label{fig6}
\end{figure}

In Fig. 5 experimental data are shown as solid circles. The solid line represents results of application of the scaling law developed here. The calculation was performed for a cloud with $r_l=60$; $n_0=0.05$. The total number of atoms was $N=6800$. The transverse size and corresponding optical depth were changed in such a way that $N$ was fixed. Then the calculated intensity was modified according to relation (\ref{6}) to achieve the experimental parameters.  Note that for large transverse sizes the aspect ratio for a Gaussian cloud in the calculation  becomes essentially different from the experimental one. It restricts the possibility to consider clouds with small optical depth in the present theory.

\section{Conclusions}
In this paper, a microscopic quantum approach to calculation of the differential light scattering cross section has been combined with numerical simulations to study the size and density scaling of the cross-section.  Particular attention has been paid to the case of a cold and dense, but not quantum degenerate atomic gas.  The principal result of this paper is the recognition of a scaling law for the scattering cross section differential in angle and in the frequency of the incident light. Comparison of this result with available experimental data is also made, and quite satisfactory agreement is found over a substantial range of resonant optical depths.  As we have explored here, the limitations of the developed scaling law as a function of scattering angle are seen to be mainly due to coherent effects such as the size of the coherent backscattering cone and scattering into the diffractive zone.

We expect that this result will permit further analysis of light scattering from larger and dense atomic samples, and in a variety of angular configurations.  However, we should point out that though such a semiquantitative estimation as in Eq. (\ref{6}) is made on the basis of the observations of the results of numerical simulations over a quite wide range of conditions,  we do not expect such a simple scaling of the light scattering cross section to apply to the situation where the dominant part of observed signals is contributed by localized configurations of atoms and light.

\subsection{Acknowledgments}
This work was supported by the Russian Foundation for Basic Research (RFBR grant No. 10-02-00103), (RFBR-CNRS grant 12-02-91056), and by the National Science Foundation (Grant No. NSF-PHY-1068159). A.S.K. thanks the Fund of Non-Profit Programs "Dynasty".

\end{document}